\documentclass[12pt]{article}
\usepackage{setspace}
\usepackage{xspace, amsmath, amssymb}

\author{\normalsize Jun Chul Park \footnote{E-mail address: junchul@kaist.ac.kr}\\
{\em \small{Department of Physics, KAIST, Daejeon 305-701,
Korea}}}

\title{\bf Representation of intermediate time-scale motions in stochastic
modeling:\\ \large Analysis on stochastic description of classical
Hamiltonian dynamics in relation with measurement imperfection}

\date{}

\begin{document}
\maketitle

\begin{abstract}
It is a well established result that, in classical dynamical systems
with sufficient time-scale separation, the fast chaotic degrees of
freedom are well modeled by (Gaussian) white noise. In this paper,
we present the stochastic dynamical description for intermediate
time-scale motions with insufficient time-scale separation from the
slow dynamical system. First, we analyze how the fast deterministic
dynamics can be viewed as stochastic dynamics under experimental
observation by intrinsic errors of measurement. Then, we present how
the stochastic dynamical description should be modified if
intermediate time-scale motions exist: the time correlation of the
noise $\xi$ is modified to $\left < \xi(t)\xi(t') \right > =
\mathcal{C}(x,p)\delta (t-t')$, where $\mathcal{C}(x,p)$ is a smooth
function of the slow coordinate $(x,p)$, and generally the cumulants
of $\xi$ except its average vary as a smooth function of the slow
coordinates $(x,p)$. The analysis given in this work actually shows
that, regardless of the sufficiency of time-scale separation, any
complex (chaotic and ergodic) dynamical system can be well described
using Markov process, if we perfectly construct the deterministic
part of (extended) stochastic dynamics.
\end{abstract}

\section{Introduction}
In classical many-particle systems with time-scale separation such
as Brownian motion, the rapid dynamic fluctuations by numbers of
fast and chaotic degrees of freedom are well modeled by Gaussian
white noise, and the Langevin equation gives successful descriptions
for such thermodynamic systems \cite{Zwanzig,Kubo1966}. Moreover,
recent theoretical works
\cite{Riegert2005,Just2001,Jarzynski1995,Kolovsky1994} indicate
that, even for few fast chaotic degrees of freedom, the dynamic
fluctuations are well approximated by suitable stochastic processes
with white noise and the stochastic dynamics still gives reasonable
descriptions. Such results certify that the stochastic process is
generally a well founded representation for fast chaotic degrees of
freedom, if the time-scale separation is sufficient appropriately.
Actually the stochastic modeling of fast chaotic dynamics based on
time-scale separation has been applied successfully in diverse areas
of science such as hydrodynamics \cite{Graham1973,Haken1975},
climate models \cite{Majda1999}, and chemical and biological systems
\cite{Haken1975,Cross1993}.  In many cases of real physical systems,
however, the corresponding dynamical systems contain diverse
\textit{intermediate} time-scale motions and cannot be well
decomposed or approximated simply by two sub-systems with the
sufficient time-scale separation. In this paper, we address the
problem of how the intermediate time-scale motions should be
represented in the point of view of stochastic dynamics. We clarify
the role of the intermediate time-scale dynamics in the stochastic
description of classical Hamiltonian systems and its qualitative
difference from the fast and slow dynamics which are represented as
Gaussian white noise and the deterministic part in the Langevin
equation, respectively.

Consider the Hamiltonian
\begin{equation}
\mathcal{H}=\mathcal{H}_{slow}(x,p)+\mathcal{H}_{fast}(q_i,p_i)+V(x,q_i),
\end{equation}
in which a slow system $\vec{x}_s\equiv(x,p)$ is interconnected with
a fast system
$\vec{x}_f\equiv(q_i,p_i)\equiv(q_1,p_1,\dots,q_N,p_N)$ by a
potential $V(x,q_i)$. Assuming that the time-scale separation is
sufficient, for the motion of $p$ by $\mathcal{H}_{slow}$, the
potential $V(x,q_i)$ introduces additional dynamics, which is
composed of slow dynamics, such as damping in Brownian motion
\cite{Berry1993}, and rapid fluctuating dynamics. We represent the
rapid fluctuating term in the equation of motion for $p$ as
$k(x,p,q_i,p_i)$ and write
$\frac{dp}{dt}=-\partial_{x}\mathcal{H}_{slow}-\partial_{x}V=h(x,p)+k(x,p,q_i,p_i)$,
in which the slow dynamics by $V$ is contained in
$h(x,p)$.\footnote{As is well known from the projection operator
method \cite{Zwanzig}, in equilibrium thermodynamical system,
$h(x,p)$ can be extracted by the projection operation, if one can
perfectly construct the relevant subspace in Hilbert space.}  Then,
we consider the following deterministic system:
\begin{equation}
\begin{aligned}
&\frac{dx}{dt}=p,
&&\frac{dp}{dt}=h(x,p)+ k(\vec{x}),\\
&\frac{dq_i}{dt}=u_i(\vec{x}), \qquad
&&\frac{dp_i}{dt}=v_i(\vec{x}), \label{1}
\end{aligned}
\end{equation}
where $\vec{x}\equiv (x,p,q_1,p_1,\dots,q_N,p_N)\equiv
(\vec{x}_s,\vec{x}_f)$ and $i=1,\dots,N$. In the case that we cannot
know or it is unnecessary to know the exact information for the fast
variable part in the equations of motion (\ref{1}), we may
approximate (\ref{1}) to stochastic differential equations, for
example,
\begin{equation}
\frac{dx}{dt}=p, \qquad \frac{dp}{dt}=h(x,p)+ \xi, \label{2}
\end{equation}
with a suitable time correlation function
$\left<\xi(t)\xi(t')\right>$ and a Gaussian probability density
function (PDF) for a random variable $\xi$.
 In this case, the stochasticity in (\ref{2}) is just a result of mathematical
approximation or simplification of the original complex system. Such
approximation can be validated only through the observed empirical
data for the system (\ref{1}). Here, it should be noted that
experimental data always contains unpredictable intrinsic errors due
to the imperfection of measurement process, which destroy the
information of fine dynamical structures. The `unpredictable' nature
of error makes such fine dynamics change into stochastic dynamics in
the resultant experimental data. If we assume all obtainable
deterministic information for (1) in our measurement systems is
described by the term $h(x,p)$, actually the mathematically
introduced stochasticity by $\xi$ should correspond to the
inevitable stochasticity in the experimental data originated by the
intrinsic measurement errors: the stochasticity contained in the
experimental data gives the physical identification for the
mathematically introduced.  In what follows, we investigate how the
stochasticity originated by time measurement errors is related with
the fine dynamical structures.
\section{Stochasticity induced by time measurement error}
Let us assume that there are inherent errors in our measurement
system for coordinates and time. We obtain an erroneous value
$(\vec{x},t)$ in measurement at an arbitrary instant, while the
\textit{ideal coordinates} and \textit{ideal time}, which is
measured by a perfect measurement system without error at the same
instant, is $(\vec{x}^*,t^*)=(\vec{x},t)+(\Delta \vec{x},\Delta t)$;
that is, $(\Delta \vec{x},\Delta t)$ is the error in our
measurement. Under our measurement system, the dynamics of (\ref{1})
is observed as the time series $\vec{x}(t)$, which is deformed from
the original dynamics of (\ref{1}), as follows:
\begin{equation}
\vec{x}(t)=\vec{x}^*(t^*) + \Delta \vec{x},
\end{equation}
where $\vec{x}^*(t^*)$ satisfies the equation of motion (\ref{1})
exactly.   In this work, we analyze how the dynamics of $\vec{x}(t)$
changes as the accuracy of our measurement system changes;
%
if we assume that, for an arbitrary measurement system, the
statistical properties of the error $(\Delta \vec{x},\Delta t)$ is
always independent of time and the error occurrences at different
times are probabilistically independent of each other, the
time-dependence of $\vec{x}(t)$ only comes from $\vec{x}^*(t^*)$,
and the coordinate measurement error $\Delta \vec{x}$ does not have
any role in our analysis---only the time measurement error $\Delta
t$ is important in analyzing the variation of the \textit{dynamical}
behavior of $\vec{x}(t)$ in relation with the change of the accuracy
of measurement system.\footnote{In this respect, time measurement
error plays a special role in dynamics.} As can be easily checked,
all results given in this work essentially hold, regardless of the
presence of the effect by coordinates measurement error $\Delta
\vec{x}$. Thus, to avoid notational complexes and to simplify the
arguments, we only consider the effect by time measurement error
$\Delta t$, which does not make any loss of the generality of the
analysis.  We assume that, except time, we are perfectly informed
about all values of the observables $x$, $p$, $q_i$, and $p_i$ in
(\ref{1}) at each instant from a perfect observer.  The situation
under our consideration is that there is a missed
time-parametrization for the perfectly observed $\vec{x}$ at each
instant, i.e.,
\begin{equation}
\vec{x}(t)=\vec{x}^*(t^*),
\end{equation}
but generally
\begin{equation}
t\not=t^*.
\end{equation}
In the following, we use the notation for $\vec{x^*}(t^*)$ as
\begin{equation}
\vec{x}^*(t^*)\equiv\vec{x}(t^*)
\end{equation}
in order to emphasize that $\vec{x}=\vec{x}^*$ at an arbitrary
instant; we obtain the time series $\vec{x}(t)$, while the perfect
observer obtains $\vec{x}(t^*)$, which satisfies the equations of
motion (\ref{1}) exactly.

%
%
%

We can characterize the time measurement errors by giving the PDF
$f_{t}(\Delta t)$, for which the probability to obtain an error in
the range $(\Delta t,\Delta t+d(\Delta t))$ is given as
$f_{t}(\Delta t)d(\Delta t)$. Then, we can be confident
about the measured data $t$ within an error range as 
\begin{equation}
t^*-\frac{\varepsilon}{2}<t<t^*+\frac{\varepsilon}{2},
\end{equation}
where we can define the suitable $\frac{\varepsilon}{2}$ based on
the PDF $f_{t}(\Delta t)$. Thus, for two measured times $t_1$ and
$t_2$ such that $0<t_2-t_1<\varepsilon$, we cannot be certain
$t^*_2>t^*_1$ because their error-ranges are overlapped, i.e., to
ensure that the measured times give the correct time order of
$t^*_1$ and $t^*_2$, the measured times should be separated as
$|t_2-t_1|\geq \varepsilon$.  In other words, approximately there is
a \textit{minimum length of time elapse} that can be identified by a
given time measurement system (one cannot chase correctly the
causality of the events occurring within a time interval shorter
than the minimum), and this minimum length of time elapse is
determined by the inherent errors in the measurement system. In
translating the deterministic system (\ref{1}) into a stochastic
system like (\ref{2}) to explain experimental data, an important
physical meaning is added to the differential $dt$. The physical
meaning of $dt$ in (\ref{2}) is the minimum length of time elapse
identifiable in our measurement system. We can determine the minimum
length of time elapse $dt$ to be $\varepsilon$ providing that the
probability to obtain a time value outside of the range
$t^*-\frac{\varepsilon}{2}<t<t^*+\frac{\varepsilon}{2}$ is very
small:
\begin{equation}
dt=\varepsilon.\label{3}
\end{equation}

Let us assume that the dynamical system (\ref{1}) starts from an
initial value of $\vec{x}$. Then, if we measure an observable
$\mathcal{A}(\vec{x})$ at $t$ for system (\ref{1}), the measured
value of $\mathcal{A}$ at $t$, i.e., $\mathcal{A}(t)$ is exactly
given by the equations of motion (\ref{1}) as
$\mathcal{A}(t^*)\equiv \mathcal{A}(\vec{x}(t^*))$, where $t^*$ is a
value in $t-\frac{\varepsilon}{2}<t^*<t+\frac{\varepsilon}{2}$:
$t^*$ randomly has a value in the range with the PDF $f_{t}(\Delta
t)$.
Thus, $\mathcal{A}(t)$ is given by one of the elements in the
following set:
\begin{equation}
\mathfrak{S}_{\mathcal{A}}(t,\varepsilon)=
\left\{\mathcal{A}(t^*)\left | \;\;
t-\frac{\varepsilon}{2}<t^*<t+\frac{\varepsilon}{2}\right.\right\},\label{4}
\end{equation}
which is the set of all values obtainable in the measurement of
$\mathcal{A}$ at $t$. Exactly which value in
$\mathfrak{S}_{\mathcal{A}}(t,\varepsilon)$ is given for
$\mathcal{A}$ in our measurement at $t$ is totally a probabilistic
problem determined by the following two components: (i) PDF
$f_t(\Delta t)$ and (ii) the set
$\mathfrak{S}_{\mathcal{A}}(t,\varepsilon)$, where we assume that
$f_t(\Delta t)$ is invariant for time translation and the error
occurrences at different times are probabilistically independent of
each other. In what follows, we assume $f_{t}(\Delta t)$ is well
approximated as $f_{t}(\Delta t) \approx \frac{1}{\varepsilon}$ for
$-\frac{\varepsilon}{2}<\Delta t<\frac{\varepsilon}{2}$, otherwise
$f_{t}(\Delta t)\approx 0$.
\section{Conventional Langevin equation system as a special case}
Let  $\mathcal{A}$  in (\ref{4}) be the fast coordinates
$\vec{x}_f$. Then, the set
$\mathfrak{S}_{\vec{x}_f}(t,\varepsilon)\equiv
\mathfrak{S}(t,\varepsilon)$ is the trajectory of $\vec{x}_f(t^*)$
during $t-\frac{\varepsilon}{2}<t^*<t+\frac{\varepsilon}{2}$ in the
fast phase space $\Gamma$ defined by $\vec{x}_f$. As the value of
$\varepsilon$ increases, i.e., as our measurement system is less
fine, the set $\mathfrak{S}(t,\varepsilon)$ occupies larger part of
$\Omega$, where $\Omega$ is the set of all solutions $\vec{x}_f$ of
(\ref{1}) in $\Gamma$. Let us consider the \textit{extremal case}
where $\varepsilon$ is sufficiently large for $\vec{x}_f$ to wander
over almost all areas of $\Omega$ during
$t-\frac{\varepsilon}{2}<t^*<t+\frac{\varepsilon}{2}$, providing
that $\vec{x}_f$ is ergodic and chaotic {\cite{Eckmann1985}. Then,
the set $\mathfrak{S}(t,\varepsilon)$ is almost the same as $\Omega$
and independent of time, which also guarantees the time independence
of the set $\mathfrak{S}_{\mathcal{F}}(t,\varepsilon)$ for any fast
observable $\mathcal{F}(\vec{x}_f)$. Thus, if $\vec{x}_f$ is fast
enough to satisfy the \textit{extremal case} for a given
$\varepsilon$, the statistical properties of the set
$\mathfrak{S}(t,\varepsilon)$ are almost stationary for time: that
is, the statistical properties of the measured values of
$\mathcal{F}(\vec{x}_f)$ at $t$ are stationary, and
$\mathcal{F}(\vec{x}_f)$ behaves as a random variable with
stationary statistical properties. On the other hand, it should be
noted that the set $\mathfrak{S}(t,\varepsilon)$ is to be slowly
dependent on time, because $\mathfrak{S}(t,\varepsilon)$ is
determined by the fast
equations of motion from the Hamiltonian 
\begin{equation}
\mathcal{H'} \equiv \mathcal{H}_{fast}(q_i,p_i)+V(x,q_i)
\end{equation}
and the value of $\mathcal{H'}$ has a slow time-dependence through
the slow motion of $x$ originated from $h(x,p)$ in (\ref{1})
\cite{Ott1979}. Thus, the \textit{extremal case} is actually the
case where the potential $V(x,q_i)$ is regarded as an extremely slow
varying function for the deterministic slow motion of $\vec{x}_s$,
i.e., $V(x,q_i)$ is effectively invariant for the slow motion of
$\vec{x}_s$ and the trajectory of $\vec{x}_f$ is almost confined
near the hypersurface $\mathcal{H'}=const$ (oscillating around the
hypersurface).\footnote{Here, we assumed that the oscillation energy
is comparatively small to the average of $\mathcal{H'}$ during $dt$
for simplicity in argument, but, without the assumption, (\ref{9})
and (\ref{10}) can be derived in similar proof process.}

Also, in the \textit{extremal case}, by the ergodic property, the
time correlation function of any fast observable
$\mathcal{F}(t^*)\equiv\mathcal{F}\left(\vec{x}_f(t^*)\right)$
approximately satisfies
\begin{equation}
\langle\mathcal{F}(t^*_1)\mathcal{F}(t^*_2)\rangle\approx 0
\;\;\;\mbox{for}\;\;|t^*_1-t^*_2|\geq\varepsilon, \label{5}
\end{equation}
where we assume $\langle\mathcal{F}(t^*)\rangle = 0$. In our
measurement system, (concerning the time measurement errors) the
experimentally ascertained quantity as the correlation function of
$\mathcal{F}(t)$ is
\begin{eqnarray}
\langle\mathcal{F}(t_{1})\mathcal{F}(t_{2})\rangle_{exp} &\equiv&
\int_{-\frac{\varepsilon}{2}}^{\frac{\varepsilon}{2}}
\int_{-\frac{\varepsilon}{2}}^{\frac{\varepsilon}{2}}
\langle\mathcal{F}(t_{1}+\Delta t_1)\mathcal{F}(t_{2}+\Delta
t_2)\rangle \nonumber \\
& &\times f_t(\Delta t_1)f_t(\Delta t_2)d(\Delta t_1)d(\Delta t_2),
\label{6}
\end{eqnarray}
where, in the integrand, we use
$\mathcal{F}(t^*_{1,2})=\mathcal{F}(t_{1,2}+\Delta t_{1,2})$.   If
$|t_1-t_2|\geq2\varepsilon$, we obtain
$\langle\mathcal{F}(t_{1}+\Delta t_1)\mathcal{F}(t_{2}+\Delta
t_2)\rangle\approx 0$ from (\ref{5}) for any $\Delta t_1$ and
$\Delta t_2$ in the integration ranges, and
\begin{equation}
\left<\mathcal{F}(t_{1})\mathcal{F}(t_{2})\right>_{exp}\approx 0
\;\;\;\mbox{for}\;\;|t_1-t_2|\geq2\varepsilon. \label{7}
\end{equation}
Thus, the correlation actually behaves as a $\delta$-function in our
measurement system, i.e.,
$\left<\mathcal{F}(t_{1})\mathcal{F}(t_{2})\right>_{exp}\sim\delta(t_1-t_2)$.\footnote{The
time $t=0$ and $t=\varepsilon$ respectively corresponds to $t=0$ and
$t \rightarrow 0$ in the case where $t$ is continuously variable.
Since $\lim_{t \rightarrow 0}\delta(t)$ needs not to be 0, (\ref{7})
is sufficient for $\delta$-function correlation.}

Let us analyze the behavior of the rapid fluctuating time series
$k(\vec{x})$ in (\ref{1}) under our measurement system.\footnote{The
term $h(x,p)$ gives the averaged dynamics for all possible
measurement errors; the errors cannot deform this deterministic
dynamical structure. Basically, we assume that the times series
$h(x,p)$ is sufficiently slow so that $h(t^*+\varepsilon)-h(t^*)
\approx \dot{h}(t^*)\varepsilon $, and the error-induced variation
in the measured value of $h(x,p)$ is comparatively negligible to
that of $k(\vec{x})$. Thus, we analyze the fluctuating term
$k(\vec{x})$ only.} Expanding $k(\vec{x})$ for the slow coordinates
$\vec{x}_s$, we obtain $k(\vec{x})=\mathcal{F}_{0}(\vec{x}_f)+
\sum_{l,m=1}^{\infty}\mathcal{F}_{lm}(\vec{x}_f)x^l p^m$}.
%
%
As previously argued, the first term $\mathcal{F}_{0}(\vec{x}_f)$
behaves as a white noise with stationary statistical properties in
the \textit{extremal case}. The remaining part $\Delta k(\vec{x})
\equiv k(\vec{x})-\mathcal{F}_{0} (\vec{x}_f)$ also should behave as
a white noise in the \textit{extremal case}, as shown in the
following argument.  Let us denote by $\langle \cdots
\rangle_{\mathfrak{S}_{\mathcal{A}}(t,dt)}$ the average for all
elements of $\mathfrak{S}_{\mathcal{A}}(t,dt)$. Assuming that all
linear increment of $p(t)$ during $dt$ is given by the deterministic
part as $h(x,p)dt$, we request $\langle
k(\vec{x})\rangle_{\mathfrak{S}_{\vec{x}}(t,dt)}=0$, which means
that $\langle \Delta
k(\vec{x})\rangle_{\mathfrak{S}_{\vec{x}}(t,dt)}=0$, because
$\langle \mathcal{F}_{0}
(\vec{x}_f)\rangle_{\mathfrak{S}(t,dt)}=const=0$. Thus, we can
conclude that the time series $\Delta k(\vec{x})$ is composed of the
fast fluctuating motion of $\vec{x}_s$ and the motion of
$\vec{x}_f$. If we assume that the fast fluctuating motion of
$\vec{x}_s$ has the same time-scale as $\vec{x}_f$, then the motion
satisfies a $\delta$-function correlation as (\ref{7}), and $\Delta
k(\vec{x})$ also behaves as a white noise. Thus, we can replace
$k(\vec{x})$ by a single white noise $\xi$, and the equation of
motion of $p$ is observed as
\begin{subequations}\label{8}
\begin{equation}
\frac{dp}{dt}=h(x,p)+\xi,\\
\end{equation}
\begin{equation}
\left<\xi(t)\xi(t')\right>=2C\delta(t-t')
\end{equation}
\end{subequations}
in the measurement system with $dt(=\varepsilon)$, where $C$ is a
constant.  Also, it is reasonable to assume that the PDF for $\xi$,
which is determined basically by $f_t(\Delta t)$ and
$\mathfrak{S}(t,\varepsilon)\approx\Omega$, is Gaussian
\cite{Landau}: $\xi$ can be regarded as the fluctuation from an
\textit{ideal} thermal reservoir in equilibrium, because the
behavior of $\xi$ is independent from $\vec{x}_s$ and the energy
flux between the fast and slow system during $dt$ is averagely 0.

\section{Stochastic dynamical description for insufficiently time-scale separated dynamical system}

Now, we generalize the above argument. \textit{Let us denote by
$\tau$ the minimum of the time length $\delta$ for which we can
approximate $\mathfrak{S}(t,\delta)\approx\Omega$ for the fast
dynamical system}. Then, if the relative accuracy of our measurement
system, denoted by $\mathbb{M}_{old}$, to the fast dynamics is not
fine so that $dt(=\varepsilon)\geq\tau$, it corresponds to the
\textit{extremal case}.

Let us observe the dynamical system (\ref{1}) with a finer new
measurement system denoted by $\mathbb{M}_{new}$ such that
$dt<\tau$. Then, there is some non-ignorable area in $\Omega$ which
can not be reached by $\vec{x}_f$ during $dt$, and we cannot simply
approximate $\mathfrak{S}(t,dt)\approx\Omega$, i.e., we have
$\mathfrak{S}(t,dt) \subsetneq \Omega$. The set $\mathfrak{S}(t,dt)$
becomes dependent on $t$, and the statistical properties of the set
$\mathfrak{S}(t,\varepsilon)$ are generally non-stationary:
$\mathcal{F}(\vec{x}_f)$ generally behaves as a random variable with
non-stationary statistical properties. Considering
$\mathfrak{S}(t,dt)$ is the trajectory of $\vec{x}_f(t^*)$ during
$t-\frac{dt}{2}<t^*<t+\frac{dt}{2}$, its time-dependence indicates
that $\mathfrak{S}(t,dt)$ is dependent on the value of
$\vec{x}(t^*)\equiv(\vec{x}_s,\vec{x}_f)(t^*)$ at $t^*=t$. In here,
each $\vec{x}_s$-dependence and $\vec{x}_f$-dependence has some
different dynamical meaning. In the following analysis, we will show
that the $\vec{x}_f$-dependence is actually related with the
deterministic slow motions which are newly emerged in
$\mathbb{M}_{new}$ and the $\vec{x}_s$-dependence is related with
intermediate time-scale motions.

If we expand $k(\vec{x})$ in (\ref{1}) for the fast coordinates
$\vec{x}_f$, we obtain $k(\vec{x})=\sum_{|\alpha|\geq 0}
\mathcal{S}_{\alpha}(\vec{x}_s) {\vec{x}_f}^{\alpha}$, where
$\alpha=(\alpha_1,\dots,\alpha_{2N})$ is a multi-index and we can
set $\mathcal{S}_{0}(\vec{x}_s)=0$. Then, the linear term
$\tilde{k}(\vec{x}_f) \equiv \sum_{|\alpha|=1}
\mathcal{S}_{\alpha}(0)\vec{x}_f^{\alpha}$ generates the most slow
dynamics in $k(\vec{x})$---while all ${\vec{x}_f}^{\alpha}$ for
$|\alpha|\geq 1$ are detected as white noises in $\mathbb{M}_{old}$,
$q_i$ and $p_i$ generate the most slow time series among
${\vec{x}_f}^{\alpha}$. As our measurement system changes from
$\mathbb{M}_{old}$ to $\mathbb{M}_{new}$, the variation of the
statistical properties (the moments or cumulants) of the measured
values of $k(\vec{x})$ at $t$ mainly comes from the most slow time
series $\tilde{k}(\vec{x}_f)$.
In the following, we analyze how the most slow time series
$\tilde{k}(\vec{x}_f)$ is viewed in $\mathbb{M}_{new}$ ($dt<\tau$).

\subsection{$\vec{x}_f$-dependence of the set
$\mathfrak{S}(t,dt)$} Let us assume that $\mathfrak{S}(t,dt)$ is
only dependent on $\vec{x}_f$ and almost independent of $\vec{x}_s$.
As previously argued in the \textit{extremal case}, the
$\vec{x}_s$-independence of $\mathfrak{S}(t,dt)$ actually means that
the motion of $\vec{x}_f$ is nearly confined on a hypersurface
$\mathcal{H'}=const$;  $\Omega$ is approximately given as the
$\mathcal{H'}=const$ surface. Thus, in this case, the
$\vec{x}_f$-dependence means that $dt$ is not long enough for
$\vec{x}_f$ to wander over all areas of the $\mathcal{H'}=const$
surface.  Consider the averaged dynamics $\langle \vec{x}_f
\rangle_{\mathfrak{S}(t,dt)}$. Because $\vec{x}_f$ cannot wander
over all areas of the $\mathcal{H'}=const$ surface during $dt<\tau$,
generally $\langle\vec{x}_f\rangle_{\mathfrak{S}(t,dt)} \neq const$
for time in $\mathbb{M}_{new}$; on the contrary, in
$\mathbb{M}_{old}$ with $dt\geq\tau$, $\langle \vec{x}_f
\rangle_{\mathfrak{S}(t,dt)}=const=0$ (with a suitable setting of
the origin of $\Gamma$), because $\mathfrak{S}(t,dt)\approx
\mathfrak{S}(t,\tau)\approx \Omega$ from the definition of $\tau$.
In $\mathbb{M}_{old}$ with $dt\geq\tau$, $\langle
\vec{x}_f\rangle_{\mathfrak{S}(t,dt)}$ is stationary as is fixed at
the origin of $\Gamma$ but, as our measurement system is finer so
that $dt<\tau$, $\langle \vec{x}_f \rangle_{\mathfrak{S}(t,dt)}$ is
not fixed any more and starts to move slowly with time in
$\Gamma$.\footnote{$\langle \vec{x}_f\rangle_{\mathfrak{S}(t,dt)}$
is a function $\vec{x}_f$ from the $\vec{x}_f$-dependence of
$\mathfrak{S}(t,dt)$.}
That is, the ensemble average of the measured values of $\vec{x}_f$
at each instant varies slowly over time in $\mathbb{M}_{new}$, which
is an undetectable deterministic motion in $\mathbb{M}_{old}$ and
viewed as a stochastic process in $\mathbb{M}_{old}$.  Considering
$\langle \tilde{k}(\vec{x}_f)
\rangle_{\mathfrak{S}(t,dt)}=\sum_{|\alpha|=1}
\mathcal{S}_{\alpha}(0)\langle \vec{x}_f^{\alpha}
\rangle_{\mathfrak{S}(t,dt)}$ also slowly varies by the motion of
$\langle \vec{x}_f \rangle_{\mathfrak{S}(t,dt)}$, eventually, this
newly emerged slow dynamics induces the observer using
$\mathbb{M}_{new}$ to change the deterministic term $h(x,p)$ in
(\ref{2}) to $h(x,p)+\delta h(x,p)$ by adding the more detailed
information of the dynamics.

On the other hand, the motion of the fast fluctuating part
%
\begin{equation}
\vec{y}_f\equiv\vec{x}_f-\langle
\vec{x}_f\rangle_{\mathfrak{S}(t,dt)}
\end{equation}
is confined on the energy surface $\mathcal{H'}=const$ in the
$\vec{y}_f$ phase space (note that $\langle
\vec{x}_f\rangle_{\mathfrak{S}(t,dt)}$ is a function of
$\vec{x}_f$), and the set $\mathfrak{S}_{\vec{y}_f}(t,dt)$ contains
all points of the $\mathcal{H'}=const$ surface, under the assumption
that the motion of $\vec{y}_f$ is still ergodic and chaotic: if
$\vec{y}_f$ cannot wander over all areas of the energy surface
during $dt$, generally $\langle
\vec{y}_f\rangle_{\mathfrak{S}(t,dt)} \neq const$, which contradicts
with the definition of $\vec{y}_f$. Thus, the motion of $\vec{y}_f$
becomes the \textit{extremal case} again in $\mathbb{M}_{new}$ with
$dt<\tau$.

There is just a replacement of the deterministic part $h(x,p)$ by
$h(x,p)+\delta h(x,p)$ \footnote{Strictly speaking, $\delta h(x,p)$ is over-simplified expression for the newly emerged slow dynamics.  If we define a new slow variable $x'\equiv \langle \tilde{k}(\vec{x}_f)
\rangle_{\mathfrak{S}(t,dt)}$, then $\frac{dp}{dt}=h(x,p)+x'+\sum_{|\alpha|=1}
\mathcal{S}_{\alpha}(0)\vec{y}_f^{\alpha}$. Generally, $x'$ may be independent of $x$ and $p$ in the sense that, for some $t_1 \not = t_2$, $(x,p)(t_1)=(x,p)(t_2)$ but $x'(t_1) \not = x'(t_2)$. We have to contain the new additional equation of motion in the deterministic part as (in the most general expression) $\frac{dx'}{dt}=h_{new}(x,p,x',t)$. But, $x'(t)$ is very slow varying time series, and there can be one-to-one correspondence between $(x,p,x')$ and $t$ for very long time interval. Thus, we can write the deterministic part of the equations of motion as  $\frac{dx}{dt}=p$,  $\frac{dx'}{dt}=h_{new}(x,p,x')$,  and $\frac{dp}{dt}=h(x,p)+x'$ for very long time. Under $\mathbb{M}_{new}$, generally the number of the slow variables may be increased, e.g., $\vec{x}_s=(x,p,x')$; thus, in our notation, we can regard $(x,p)$ under $\mathbb{M}_{new}$ as a more expanded slow variable set than $(x,p)$ under $\mathbb{M}_{old}$. If we intend to write the equations of motion only using $x$ and $p$ except $x'$, generally the equations of motion have to contain the \emph{memory term} for $(x,p)$ instead of $x'$.} and the fluctuating noise part is identical
to the case of the Langevin equation or the \textit{extremal case},
i.e., white noise with stationary statistical properties; the
$\vec{x}_f$-dependence of the set $\mathfrak{S}(t,dt)$ changes the
time series $\tilde{k}(\vec{x}_f)$ from white noises in
$\mathbb{M}_{old}$ to colored noises in $\mathbb{M}_{new}$ as the
combination of the deterministic motion $\delta h(x,p)$ and white
noises.

\subsection{$\vec{x}_s$-dependence of the set
$\mathfrak{S}(t,dt)$} Next, we analyze the $\vec{x}_s$-dependence of
the set $\mathfrak{S}(t,dt)$. Let us assume that
$\mathfrak{S}(t,dt)$ is only dependent on $\vec{x}_s$ and almost
independent of $\vec{x}_f$; the $V(x,q_i)$ is not effectively
invariant for the slow motion of $\vec{x}_s$ any more.  Actually,
the $\vec{x}_f$-independence means $\langle
\vec{x}_f\rangle_{\mathfrak{S}(t,dt)}=const=0$
for time, as shown in the following argument. If we fix the value of
$\vec{x}_s$ arbitrarily, the set $\mathfrak{S}(t,dt)$ is invariant
for any value of $\vec{x}_f$ on the hypersurface
$\mathcal{H'}=const$ corresponding to the fixed value of $\vec{x}_s$
(note that $\mathcal{H}$ is conserved), which means that the set
$\mathfrak{S}(t,dt)$ contains all points $\vec{x}_f$ on the
hypersurface. Also, since $\mathcal{H'}$ varies slowly with the slow
motion of $\vec{x}_s$, the elements of $\mathfrak{S}(t,dt)$ are
confined near the hypersurface. Thus, actually the average for the
set $\mathfrak{S}(t,dt)$ can be approximated as the average for all
points on the $\mathcal{H'}=const$ surface and $\langle
\vec{x}_f\rangle_{\mathfrak{S}(t,dt)}=0$, which holds for arbitrary
values of $\vec{x}_s$.

The $\vec{x}_s$-dependence induces the change of the set
$\mathfrak{S}(t,dt)$ while keeping $\langle
\vec{x}_f\rangle_{\mathfrak{S}(t,dt)}=0$ invariant for time, which
means the following two results:
\begin{itemize}
\item [(i)]the variation of $\vec{x}_s$ induces the variation of the
other statistical properties of $\mathfrak{S}(t,dt)$, such as the
averages of $q_i^2$, $q_ip_j$, $q_iq_jp_k$, $\dots$, which are
stationary in $\mathbb{M}_{old}$, i.e., generally the statistical
properties $\langle \vec{x}_f^{\alpha} \rangle_{\mathfrak{S}(t,dt)}$
for $|\alpha|\geq2$ are non-stationary and vary as a smooth function
of $\vec{x}_s$ in $\mathbb{M}_{new}$,\footnote{The smoothness of
$\langle \vec{x}_f^{\alpha} \rangle_{\mathfrak{S}(t,dt)}$ as a
function of $\vec{x}_s$ comes from the fact that $\lim_{t_1
\rightarrow t_2}\mathfrak{S}(t_1,dt)=\mathfrak{S}(t_2,dt)$ and the
trajectory of $\vec{x}_f$ in $\Gamma$ is smooth.} and
\item [(ii)] the $\vec{x}_s$-dependence does not make any change
in the deterministic part $h(x,p)$ in (\ref{2}).
\end{itemize}
Except $\langle \tilde{k}(\vec{x}_f) \rangle_{\mathfrak{S}(t,dt)}$,
generally $\langle
\tilde{k}^{\alpha}(\vec{x}_f)\rangle_{\mathfrak{S}(t,dt)}$ for
$|\alpha| \geq 2$ vary as a function of $\vec{x}_s$ in
$\mathbb{M}_{new}$, i.e.,
the statistical properties of the measured values of
$\tilde{k}(\vec{x}_f)$ at $t$, except the ensemble average of
$\tilde{k}(\vec{x}_f)$, vary slowly as a function of $\vec{x}_s$ in
$\mathbb{M}_{new}$.

Also, the time correlation function of $\tilde{k}(t^*)$ is given as
$\langle \tilde{k}(t^*_1)\tilde{k}(t^*_2)\rangle\approx 0$ for
$|t^*_1-t^*_2| \geq dt$, because $dt$ is sufficient for $\vec{x}_f$
to wander over all areas of an energy surface---the
$\mathfrak{S}(t,dt)$ contains all $\vec{x}_f$ on the energy surface.
Thus, considering $\langle \tilde{k}^2(\vec{x}_f)
\rangle_{\mathfrak{S}(t,dt)}$ should be a smooth function of
$\vec{x}_s$ and using (\ref{6}) and (\ref{7}), the motions inducing
the $\vec{x}_s$-dependence give the following time correlation for
$\tilde{k}(t)$ in $\mathbb{M}_{new}$ with $dt<\tau$:
\begin{equation}
\left<\xi(t_1)\xi(t_2)\right>_{exp}=\mathcal{C}(x,p)\delta(t_1-t_2),
\label{9}
\end{equation}
where the notation $\tilde{k}(\vec{x}_f)$ is replaced by $\xi$ and
$\mathcal{C}(x,p)$ is a smooth function of $\vec{x}_s$. Therefore,
under the $\vec{x}_s$-dependence, $\tilde{k}(\vec{x}_f)$ still
behaves as a white noise as satisfies (\ref{9}), but, differently
from the conventional white noise with stationary statistical
properties, its moments or cumulants except its average are
non-stationary for time as a smooth function of the slow coordinates
$\vec{x}_s$.

\subsection{Extension in Stochastic dynamical formalism}
As is clear from the above analysis, the motions of $\vec{x}_f$
generating the $\vec{x}_s$-dependence have a faster time-scale than
the motions generating the $\vec{x}_f$-dependence which are observed
as the deterministic slow dynamics $\delta h(x,p)$: under the
$\vec{x}_f$-dependence, $\vec{x}_f$ is not fast enough to wander
over all areas of an energy surface during $dt$ but, under the
$\vec{x}_s$-dependence, $\vec{x}_f$ is fast enough to do so. Also,
the motions of $\vec{x}_f$ generating the $\vec{x}_s$-dependence
have a slower time-scale than the motions of $\vec{y}_f \equiv
\vec{x}_f-\langle \vec{x}_f\rangle_{\mathfrak{S}(t,dt)}$ in the case
of $\vec{x}_f$-dependence, i.e., the fast system in the
\textit{extremal case} (the fast system described as white noise in
the Langevin equation): under the $\vec{x}_s$-dependence, the
motions of $\vec{x}_f$ cannot be approximated as confined near an
energy surface,
and, since the elements of $\mathfrak{S}(t,dt)$ are confined near an
energy surface, $\vec{x}_f$ cannot wander over the entire areas of
$\Omega$ during $dt$.  Thus, actually the motions related with the
$\vec{x}_s$-dependence are intermediate time-scale motions.

Consequently, the previously obtained results for the case of
$\vec{x}_s$-dependence show that, if there exist intermediate
time-scale motions, generally the fast dynamics, which cannot be
written in the deterministic part of stochastic dynamics, should be
described as the white noise $\xi$ of which the cumulant
\begin{equation}
c_n\equiv \left [
(-i)^n\frac{d^n}{dk^n}\ln\sum_{m=0}^{\infty}\frac{1}{m!}\left<\xi^m
\right>(ik)^m \right ]_{k=0}
\end{equation}
for $n\geq 2$ is a smooth function of the slow coordinates
$\vec{x}_s$.\footnote{The PDF for $\xi$ is a function of
$\vec{x}_s$, and generally the functional form of the PDF varies as
$\vec{x}_s$ varies.}
Thus, we have to treat the cumulants $c_n(x,p)$ as dynamical
variables in stochastic dynamics and have the following stochastic
dynamics for the coordinates $(x,p,c_2,c_3,\dots)$ with the setting
$c_1=0$:
\begin{subequations}\label{10}
\begin{equation}
\frac{dx}{dt}=p, \qquad \frac{d p}{dt}=h(x,p)+\xi,
\end{equation}
\begin{equation}
\frac{dc_n}{dt}=\frac{\partial c_n}{\partial
x}\frac{dx}{dt}+\frac{\partial c_n}{\partial p}\frac{dp}{dt}\equiv
h_n(x,p,\xi),
\end{equation}
\end{subequations}
where $n= 2, 3,\dots$, and
\begin{equation}
\left<\xi(t)\xi(t')\right>=c_2(x,p)\delta(t-t') \label{10-2}
\end{equation}
by the result (\ref{9}); the cumulants $c_n$ and (\ref{10-2})
completely determine any time correlations
$\left<\xi^{n_1}(t_1)\cdots\xi^{n_l}(t_l)\right>$, where
$n_1,\dots,n_l$ are positive integers, and thus the formulation
(\ref{10}) together with (\ref{10-2}) gives a complete stochastic
dynamical description.   In deriving (\ref{10}), similarly as in the
\textit{extremal case}, $k(\vec{x})$ can be replaced by a single
noise $\xi$: assuming that all linear increment of $p(t)$ during
$dt$ is given by the deterministic term as $h(x,p)dt$, we have
$\langle \tilde{\Delta}
k(\vec{x})\rangle_{\mathfrak{S}_{\vec{x}}(t,dt)}=0$, where
$\tilde{\Delta} k(\vec{x}) \equiv k(\vec{x})-\tilde{k}(\vec{x}_f)$,
and, also if we assume that the fast fluctuation of $\vec{x}_s$ has
the same time-scale as $\vec{x}_f$, $\tilde{\Delta} k(\vec{x})$
should satisfy at least (\ref{9}). However, generally the time-scale
of $\vec{x}_f^{\alpha}$ becomes faster as $|\alpha|$ increases, and,
for some $|\alpha|=M$, the term $\sum_{|\alpha|\geq M}
\mathcal{S}_{\alpha}(\vec{x}_s) \vec{x}_f^{\alpha}$ behaves as a
white noise. Thus, more precisely, we can write as
$\frac{dp}{dt}=h(x,p)+\xi+\eta$, where $\eta$ is the conventional
Gaussian white noise and $\langle \xi(t)\eta(t') \rangle \propto
\delta(t-t')$. While the trivial case ($c_2=const \not =0$ and
$c_n=0$ for $n\geq 3$) of formulation (\ref{10}) gives the
conventional Langevin equation, in the most simple nontrivial case
as all $c_n=0$ except $c_2 \not = const$, (as shown in the
subsequent argument) the formulation can be reduced to the
conventional multiplicative noise method.

The results obtained so far can be summarized as follows.
Intermediate time-scale motions cannot be described by deterministic
trajectories in phase space, because they satisfy a
$\delta$-function time correlation as (\ref{9}), and also they
cannot be described simply by random forces acting on the
trajectories, because they have deterministic properties represented
by the cumulants $c_n(x,p)$ for $n \geq 2$. Intermediate time-scale
motions should be described by the trajectories in the
\textit{cumulant space} defined by the coordinates $(c_2,
c_3,\dots)$. This indicates that the simple trajectory-based
description in phase space, such as the Langevin or generalized
Langevin equation, cannot describe systematically the dynamical
effect from diverse intermediate time-scale motions. Also, since
intermediate time-scale motions satisfy a $\delta$-function time
correlation, it means that, \textit{regardless of the sufficiency of
time-scale separation, any complex (chaotic and ergodic) dynamical
system can be well described using Markov process, if we perfectly
construct the deterministic part in stochastic dynamics}.

In formulation (\ref{10}), $\xi$ and $(x,p)$ exchange their
influence with each other. Thus, the random noise $\xi$ cannot be
considered as the thermal fluctuation from an \textit{ideal} thermal
reservoir in that the statistical properties of $\xi$ are not
independent from the system described by $(x,p)$---the thermal
reservoir by the fast chaotic dynamics of $\vec{x}_f$ is treated as
a \textit{finite} thermal system not having infinite capacity.

In the most simple case of formulation (\ref{10}) as all $c_n=0$
except $c_2=2\mathcal{C}(x,p)$, we have $\langle \xi(t)\xi(t')
\rangle=2\mathcal{C}(x,p)\delta(t-t')$ from (\ref{10-2}), and the
PDF $w(\xi)$ for $\xi$ is given as a Gaussian: $w(\xi) \propto
e^{-\xi^2/4\mathcal{C}(x,p)}$. If we define $\zeta \equiv
\xi/\mathcal{C}(x,p)^{\frac{1}{2}}$, the variance of the random
variable $\zeta$ is constant, and, using $\zeta$, we can reformulate
the most simple case of (\ref{10}) in a conventional multiplicative
noise system, as follows:
\begin{subequations}
\begin{equation}
\frac{d p}{dt}=h(x,p)-\frac{1}{2}\frac{\partial
\mathcal{C}(x,p)}{\partial p}
+\mathcal{C}(x,p)^{\frac{1}{2}}\zeta,\\
\end{equation}
\begin{equation}
\langle\zeta(t)\zeta(t')\rangle=2\delta(t-t'), \label{10-1}
\end{equation}
\end{subequations}
where the equation of motion for $x$ is the same as in (\ref{10}),
and the PDF for $\zeta$ is Gaussian: $w(\zeta) \propto
e^{-\zeta^2/4}$. As can be easily checked, these two systems give a
same Fokker-Planck equation; in this respect, the multiplicative
noise method is equivalent to the most simple case of formulation
(\ref{10}).

\section{Additional remarks}
Finally, we point out that the arguments for intrinsic time
measurement errors symmetrically hold for space measurement errors
in static problems, e.g., $\partial_x\psi(x,t)=
h_{s}(x,t)+h_{f}(x,t)$ where $h_{s}$ and $h_{f}$ are slow and fast
varying parts for $x$, respectively. There is the experimentally
identifiable minimum distance $dx$ corresponding to $dt$ (in
dynamics), and we obtain the stochastic statics,
$\partial_x\psi(x,t)= h_{s}(x,t)+\xi(x)$ for a random variable
$\xi(x)$. The stochasticity originated by space measurement errors
reflects the fine structure of statics on the statistical properties
of $\xi(x)$.

\section*{\normalsize Acknowledgements}
The author would like to thank Prof. Hae Yong Park, Prof. Jae-Eun
Kim and Prof. Jong-Jean Kim for their helpful discussions.

\end{document}